\documentclass[11pt]{article}
\usepackage[usenames,dvipsnames]{pstricks}
\usepackage[utf8]{inputenc}
\pdfoutput=1
\usepackage[fleqn]{amsmath}
\usepackage{amsmath,amsfonts}
\usepackage[english]{babel}
\usepackage{textcomp}
\usepackage{caption,subcaption}
\usepackage{graphicx}
\usepackage{rotating}
\usepackage{authblk}
\usepackage{epstopdf}
\usepackage{bookmark}
\usepackage{float}
\usepackage{amsmath}
\usepackage[square,numbers]{natbib}
\usepackage{color, soul}
\hypersetup{pdfauthor={some author},pdftitle={eye-catching title}}
\usepackage{ifpdf}

\title{\textbf{Radiating- collapsing models satisfying Karmarkar condition}}
\author{\small{Suresh C. Jaryal \footnote{suresh.fifthd@gmail.com}}}
\affil{\textit{Department of Physics \& Astronomical Science, }\\
\textit{ Central University of Himachal Pradesh (CUHP),\\
Dharamshala, Kangra (HP), India 176215}}
\numberwithin{equation}{section}
\date{}
\begin{document}
\maketitle
This paper presents a class of exact spherical symmetric solutions of
the Einstein equations admitting heat-conducting anisotropic fluid as a collapsing matter.
The exterior spacetime is assumed to be the Vaidya metric.
This class of solutions is shown to satisfy all the energy conditions throughout the interior 
of the star, and the luminosity is time independent, radiating 
uniformly throughout the collapse.  
\section{Introduction}
There has been extensive research in the field of 
gravitational collapse. Since the pioneering work on the gravitational collapse of homogeneous dust,\cite{OS}\cite{D},
it is now accepted that for a gravitational collapse of homogeneous pressureless matter, the central singularity remains
hidden behind the horizon, implying that the end state of the continual gravitational
collapse of homogeneous dust cloud must be a black hole \cite{Hawking_ellis}. Further studies 
have examined various aspects of gravitationally collapsing
stellar systems for different kinds of matter distributions,
and details may be found in  \cite{pankaj_book, pankaj_malafarine_review}
\cite{Santos1985}-\cite{Chan1998}.  These studies have thrown light to many interesting facts
which must hold for the collapse processes to be physically realistic.
For example, for the continuous and smooth matching of the interior collapsed
spacetime to the exterior Vaidya spacetime
over the timelike hypersurface $\Sigma$, the radial pressure must not vanish at the 
boundary of the collapsing radiant star, but instead be proportional to the heat flux\,\cite{Gauss1981}\cite{Santos}.

In order to have a physically well behaved model of a gravitation collapse with 
generic energy momentum tensors, one not only needs to find physically consistent analytical solutions 
of the Einstein field equations, but also must ensure validity of the energy conditions as well. 
The practice usually followed are: to specify the spacetime symmetry,
or the gravitational potentials, imposing an equation of state, or restricting the matter 
content to find the solutions of the gravitational collapse.
However, there exists a class $\bf{I}$ condition, which is useful to obtain classes of solutions. This condition arise
from some well known geometric theorems as follows:
First, an $(n+1)$-dimensional space 
$V^{n+1}$ can be embedded into a pseudo Euclidean space $E^{n+2}$ of dimension $(n+2)$\,\cite{Eisenhart1925a},
and that all the spherically symmetric spacetime are in general of class {\bf{II}}.
Next, the necessary and sufficient condition for any Riemannian space to be 
embedding class $\bf{I}$ is that it satisfies the Karmarkar condition \cite{Eisenhart1925b} \cite{Karmarkar}.
Thus the Karmarkar condition is a useful condition which gives new solutions.

Recently, there has been a renewed interest  
in obtaining solutions using these conditions, \cite{SP2016}-\cite{PB2017}.
The study of the non-static radiating metric with timelike Karmarkar condition, when the temporal dependence of the model is linear, shows that the collapse proceeds without the formation of the horizon\,\cite{NGM2018}.
The assumption is that the metric coefficients (which are in general, functions
of $r$ and $t$), be separable, for example $g_{rr}=b(r)f(t)$ (see the equation \eqref{1eq1} for example).
In their study, they assume one of gravitational potential, which we denote by $b(r)$, to be constant. 
Next, using the results of the horizon-free collapse model \cite{Banerjee} and  
a linear form of time function $f(t)$ into the Karmarkar condition 
the other metric function $g_{tt}$ is obtained. It must be noted that this choice of $b(r)$ is special  in \cite{NGM2018},
and one may generalise.
We introduce a form of pressure anisotropy 
$\Delta=p_t-p_r$, between the radial and the tangential pressures,
and using the Karmarkar condition,
show that here too the gravitational potentials may be determined. 
Clearly, these gravitational potentials generalise those in \,\cite{NGM2018}.
For these models, we study the thermodynamical evolution and their temperature profiles, since these quantities 
play prominent role during dissipative gravitational collapse. In fact, they decide the departure from the
thermodynamical equilibrium. Quite naturally, the study of these transport processes in 
the context of irreversible thermodynamics in radiative
gravitational collapse has gained considerable attention and
\,\cite{Maartens1995}-\cite{GG2001} give further details.

The paper has been organized as follows. In section \ref{sec2} we present a description
of field equations of spherical symmetric anisotropic system and the junction conditions for smoothly matching of the
interior spacetime with the exterior Vaidya spacetime across the timelike surface 
$\Sigma$. Also, we present the solutions to the Einstein field equation and the explicit expressions
for physical quantities. In the light of these solutions we analyze 
the physical evidences of our model by verifying the energy conditions and 
it has been shown that all these energy conditions are well satisfied throughout the collapse.
In section \ref{sec5}  we study the thermodynamical evolution and temperature
profiles of the radiative gravitational collapse. Finally, discussion of the result 
accompanied with concluding remarks are given in section \ref{sec6}.

\section{Fluid distribution and field equations}\label{sec2}
In comoving- synchronous coordinates, the general spherically symmetric metric is given by 
\begin{eqnarray}
 ds^{2}=-a(r)^2 dt^2 + b(r)^2 f(t)^2 dr^2 + r^2 b(r)^2 f(t)^2\left( d \theta^2+ \sin^2{\theta} d\phi^2 \right).
\label{1eq1}
\end{eqnarray}
We consider the interior energy-momentum tensor for the anisotropic fluid distribution 
with radial heat flow of the form
\begin{eqnarray}
T_{\mu\nu}=(p_{t}+\rho)u_\mu u_\nu +p_{t} g_{\mu\nu}+(p_{r}-p_{t})X_{\mu}X_{\nu}+q_{\mu}u_{\nu}+q_{\nu}u_{\mu}, \label{tmnz}
\end{eqnarray}
where $\rho$, $p_{r}$ and $p_{t}$ are the energy density, radial pressure and tangential pressure respectively,
$u^{\mu}$, $X^{\mu}$ and $q^{\mu}$ are unit time-like $4$-velocity vector, unit space-like vector 
along radial $4$-vector and radial heat flow vector respectively.
These satisfy $u_\mu u^\mu=-X_\mu X^\mu=-1$ and $u_\mu X^\mu=u_\mu q^\mu=0$.
In the comoving co-ordinates the $4$-velocity and unit space-like vector and radial heat flow vector of the fluid are given by 
 \begin{equation}
  \hspace{2cm} u^\mu=\frac{1}{a}\,\delta^{\mu}_0 \hspace{0.5cm};\hspace{0.5cm} X^\mu=\frac{1}{b\,f}\,\delta^{\mu}_1 
  \hspace{0.5cm} ;\hspace{0.5cm} q^\mu=\frac{1}{b\,f}\,X^{\mu} \label{uXq}.
 \end{equation}
The magnitude of the expansion scalar $\Theta$ for the metric \eqref{1eq1} is given by
 \begin{eqnarray}
\Theta&=&\bigtriangledown_{\mu}u^{\nu}=\frac{3\,\dot{f}}{a\,f}\label{Theta}.
 \end{eqnarray}
The non vanishing components of Einstein-Maxwell field equations for the 
 metric \eqref{1eq1}, energy momentum tensor \eqref{tmnz} and  \eqref{uXq}  are $($using units with $c=1=8\pi G)$ 
 \begin{eqnarray}
\rho&=&\frac{3\,\dot{f}^{2}}{a^2\,f^2}-\frac{1}{b ^{2}\,f^{2}}\left(\frac{2\,b^{''}}{b}-\frac{b^{'}\,^{2}}{b\,^{2}}+\frac{4}{r}\frac{b^{'}}{b} \right)\label{rho},\\
{p}_{r}&=&-\frac{1}{a^{2}}\left(\frac{2\,\ddot{f}}{f}+\frac{\dot{f}^{2}}{f^{2}}\right)+\frac{1}{b^{2}\,f^{2}}\left(\frac{2\,a^{'}\,b^{'}}{a\,b}+\frac{2}{r}\left(\frac{a^{'}}{a}+\frac{b^{'}}{b}\right)+\frac{b^{'}\,^{2}}{b\,^{2}} \right)
\label{pr},\\
 {p}_{t}&=&-\frac{1}{a^{2}}\left(\frac{2\,\ddot{f}}{f}+\frac{\dot{f}^{2}}{f^{2}}\right)+\frac{1}{b^{2}\,f^{2}}\left(\frac{a^{''}}{a}+\frac{1}{r}\left(\frac{a^{'}}{a}+\frac{b^{'}}{b}\right)-\frac{b^{'}\,^{2}}{b\,^{2}} +\frac{b^{''}}{b}\right)\label{pt},\\
 q&=&-\frac{2\,a^{'}\,\dot{f}}{a^{2}\,b^{2}\,f^{3}}\label{q}.
 \end{eqnarray}
where dot and prime are the derivatives with respect to $t$ and $r$ respectively.
From equations \eqref{rho}-\eqref{q} we can see that the number of field equations 
are less than that of the number of unknowns. Also, the form of three metric 
potentials fix all the unknown physical quantities of the system. So, in order to study
the collapsing phenomena we need to fix forms for these metric potentials.

Let us begin with the Israel-Darmois  junction conditions\,\cite{Darmois}-\cite{WN}. The interior manifold is $\it\bf{M^-}$
and exterior manifold is $\it\bf{M^+}$ to be matched across the bounding timelike three space $\Sigma$,
at  $r=r_b$. 
The exterior spacetime $\it\bf{M^+}$, is described by the 
Vaidya spacetime having outgoing radial flow of the radiation given by\,\cite{Vaidya}
\begin{eqnarray}
ds^{2}_{+}=-\left(1-\frac{2M(v)}{{\bf{r}}}\right)dv-2dvd{\bf{r}}+ {\bf{r}}^2 \left(d \theta^2
 +\sin^2\theta d\phi^2\right)\label{m+},
\end{eqnarray}
The junction conditions require the matching of metric as well as the extrinsic curvatures
\begin{eqnarray}
ds^2_{\Sigma}&=&(ds_{-})^2_\Sigma = (ds_{+})^2_\Sigma \label{ds}\\
\left[K_{ij}\right]_{\Sigma}&=&K_{ij}^{+}=K_{ij}^{-} \label{Kij},
\end{eqnarray}
where $K_{ij}^{\pm}$ is the extrinsic curvature to $\Sigma$ given by \cite{Eisenhart1949} 
\begin{eqnarray}
K_{ij}^{\pm}=-n_{l}^{\pm}\frac{\partial^2 x^{l} }{d z^i d z^j}-n_{l}^{\pm}\Gamma^{l}_{ mn}\frac{\partial x^{m} }
{d z^i}\frac{\partial x^{n} }{d z^j}. \label{KK}
\end{eqnarray}
Here $x^l$ are the coordinates of interior and exterior spacetimes, $z^i$ are the coordinates that defines the
 hypersurface $\Sigma$ and $n^i$ are the unit normal vector to $\Sigma$.
  %
%
The junction condition on metric functions,  given by \eqref{ds} at the hypersurface $\Sigma$ gives
\begin{eqnarray}
dt&=&a(r)_{_{\Sigma}}^{-1}\, d\tau , \label{ttau}\\
{\bf{r}}_{_{\Sigma}}(v)&=&(r\,b\,f)_{_{\Sigma}} ,\label{rR}\\ 
\left(\frac{dv}{d\tau}\right)_{\Sigma}^{-2}&=&\left( 1-\frac{2M}{\bf{r}}+2\frac{d\bf{r}}{dv}\right)_{\Sigma}, \label{vtau}
\end{eqnarray}
where $\tau$ is the time coordinate defined only on the hypersurface $\Sigma$. To 
match the extrinsic curvatures, we need the normal vector fields.
The unit normal vectors on the hypersurface $\Sigma$ for the interior and exterior spacetime are given by
\begin{eqnarray}
n^{-}_{l}&=&\left[0,(b\,f)_{_{\Sigma}},0,0\right]\label{n-},\\
n^{+}_l&=&\left(1-\frac{2M}{{\bf{r}}}+2\frac{d{\bf{r}}}{dv}\right)^{-\frac{1}{2}}_\Sigma 
\left(-\frac{d{\bf{r}}}{dv}\delta^0_l+\delta^1_l\right)_{\Sigma} \label{n+}.
\end{eqnarray}
The non vanishing components of the extrinsic curvature for metrics \eqref{1eq1} and \eqref{m+} are given by
\begin{eqnarray}
K^{-}_{\tau\tau}&=&-\left[\frac{a^{'}}{a\,b\,f}\right]_{\Sigma}, \label{Ktt-}\\
K^{-}_{\theta\theta}&=&\left[r\,b\,f\left(1+\frac{r\,b^{'}}{b} \right)\right]_\Sigma , \label{Kthth-}\\
K^{+}_{\tau\tau}&=& \left[\frac{d^2v}{d\tau^2}\left(\frac{dv}{d\tau}\right)^{-1}-\left(\frac{dv}{d\tau}\right)
\frac{M}{{\bf{r}}^2}\right]_\Sigma ,\label{Ktt+}\\
K^{+}_{\theta\theta}&=&\left[\left(\frac{dv}{d\tau}\right)\left(1-\frac{2M}{{\bf{r}}}\right)
{\bf{r}}-{\bf{r}}\frac{d{\bf{r}}}{d\tau}\right]_\Sigma ,\label{Kthth+}\\
K^{-}_{\phi\phi}&=& \sin^2{\theta} K^{-}_{\theta\theta}\,\, , \,\, K^{+}_{\phi\phi}=\sin^2{\theta}
 K^{+}_{\theta\theta} .\nonumber
\end{eqnarray}
Now, from the second junction condition \eqref{Kij}, one must have the equality for the $\theta\theta$ components
since the spherical part matches for the interior and the exterior. 
Thus, the equation $K^{+}_{\theta\theta}=K^{-}_{\theta\theta}$
 at hypersurface $\Sigma$, and along with equations \eqref{ttau}, \eqref{rR} and \eqref{vtau} give
\begin{eqnarray}
\left[r\,b\,f\left(1+\frac{r\,b^{'}}{b} \right)\right]_\Sigma&=&
\left[\left(\frac{dv}{d\tau}\right)\left(1-\frac{2M}{{\bf{r}}}\right)
{\bf{r}}-{\bf{r}}\frac{d{\bf{r}}}{d\tau}\right]_\Sigma\label{Kth+-},\\
m_{\Sigma}&=&\left[\frac{r^{3}\,\dot{f}^{2}\,b^{3}f}{2\,a^{2}}-\frac{r^{3}\,f\,b^{'}\,^{2}}{2\,b}
-r^2\,f\,b^{'}\right]_{\Sigma}, \label{mass}
\end{eqnarray}
where $2m$ is the total energy entrapped inside the hypersurface $\Sigma$\,\cite{Misner-Sharp, CM}.
Now, again from the matching condition \eqref{Kij}, the matching of the $K^{+}_{\tau\tau}=K^{-}_{\tau\tau}$ 
component together with the equations \eqref{ttau} we have
\begin{eqnarray}
-\left[\frac{a^{'}}{a\,b\,f}\right]_{\Sigma}&=&\left[\frac{d^2v}
{d\tau^2}\left(\frac{dv}{d\tau}\right)^{-1}-\left(\frac{dv}{d\tau}\right)\frac{M}{{\bf{r}}^2}\right]_\Sigma. \label{Ktt+-}
\end{eqnarray}
Substituting equations \eqref{ttau}, \eqref{rR} and \eqref{mass} into the equation \eqref{Kth+-} we have 
\begin{eqnarray}
\left(\frac{dv}{d\tau}\right)_{\Sigma}&=&\left(1+\frac{r\,b^{'}}{b}+\frac{r\,b\,\dot{f}}{a}\right)_{\Sigma}^{-1}. \label{dvdtau}
\end{eqnarray}
Now, differentiating \eqref{dvdtau} with respect to the $\tau$ and using equations 
\eqref{mass} and \eqref{dvdtau}, we can write the equation \eqref{Ktt+-} 
and comparing with equations \eqref{pr} and \eqref{q} we have the useful equation
\begin{eqnarray}
(p_{r})_{_{\Sigma}}&=&(q\,b\,f)_{_{\Sigma}}. \label{matching}
\end{eqnarray}
We also require the  total Luminosity for an observer at rest at infinity is given by\,\cite{Chan1997}
\begin{eqnarray}
L_{\infty}&=& -\left( \frac{dm}{dv}\right)_{\Sigma}=-\left[ \frac{dm}{dt}
\frac{dt}{d\tau}\left(\frac{dv}{d\tau}\right)^{-1}\right]_{\Sigma}.\label{LInf}
\end{eqnarray}
Differentiating the equation \eqref{mass} with respect to $t$ and using equation \eqref{pr}, 
\eqref{ttau} and \eqref{Ktt+-}, above equation \eqref{LInf} becomes
\begin{eqnarray}
L_{\infty}&=&\left[\frac{r^2\,f^{2}\,b^{2}\,p_{r}}{2}\left(1+\frac{r\,b^{'}}{b}+\frac{r\,b\,\dot{f}}{a}\right)^2\right]_{\Sigma}.\label{LInfinity}
\end{eqnarray}
The boundary redshift can be used to determine the time of formation of the horizon.
The boundary redshift $Z_{\Sigma}$ is given by\,\cite{Chan1997}
\begin{eqnarray}
\left(\frac{dv}{d\tau}\right)_{\Sigma}&=&1+Z_{\Sigma}. \label{red}
\end{eqnarray}
As usual, with $\left( 1+\frac{r\,b^{'}}{b}+\frac{r\,b\,\dot{f}}{a}\right)_{\Sigma}=0$, 
for an observer at rest at infinity, the red shift diverges at the time of formation of the blackhole.
Now, to find the solutions of the field equations \eqref{rho}-\eqref{q}, one need to have the form of the metric potential. 
From equations \eqref{pr}, \eqref{pt} and \eqref{1eq1}, the pressure anisotropy factor $\Delta=p_{t}-p_{r}$ has the form
\begin{eqnarray}
\Delta&=&\frac{1}{f^2\,b^2}\left[\frac{a^{''}}{a}+\frac{b^{''}}{b}
-\frac{1}{r}\left(\frac{a^{'}}{a}+\frac{b^{'}}{b}\right)-\frac{2\,a^{'}\,b^{'}}{a\,b}-\frac{2\,b^{'\,^{2}}}{b^2}
\right]. \label{Delta}
\end{eqnarray}
Since the number of field equations are more than that of the number of unknowns in the system.
These unknowns depend upon the form of metric potentials. To find the 
metric potentials, we take the pressure anisotropy $\Delta$ to be:
\begin{eqnarray}
\Delta &=& \frac{1}{f^2\,b^2}\left[\frac{a^{''}}{a}-\frac{a^{'}}{r\,a}-\frac{2\,a^{'}\,b^{'}}{a\,b}\right].\label{Delta1}
\end{eqnarray}
The full expression of anisotropy
is given in equation \eqref{Delta}. However,
this form is too complicated to be solved in
full detail. So, to simplify, we separate 
a set of terms which are significant in the following sense:
The anisotopy in pressure $\Delta$ is so assumed that the model is physically significant,
such that $\Delta$ vanishes at the center $r=0$ of the cloud and is regular towards the boundary.
Furthermore, this choice
makes the original pressure anisotropic equation \eqref{Delta} as differential equation of
only one function, which one can easily integrate to find solution. The vanishing part 
of the equation $\eqref{Delta}$
\begin{eqnarray}
0&=&\frac{1}{f^2\,b^2}\left[\frac{b^{''}}{b}-\frac{b^{'}}{r\,b}-\frac{2\,b^{'\,^{2}}}{b^2}  \right],\\
\end{eqnarray}
which gives the following solution for the function $b(r)$:
\begin{eqnarray}
b(r)&=&-\frac{2}{C_3\,r^2+2\,C_4}\label{br},
\end{eqnarray}
where $C_3$ and $C_4$ are constant of integration.

One may now obtain time dependent solutions.
It is known that an $(n+1)$-dimensional space $V^{n+1}$ can be
 embedded into a pseudo Euclidean space $E^{n+2}$ of dimension $(n+2)$ 
 if there exists a symmetric tensor $b_{\mu\nu}$ which satisfies the 
 Gauss-Codazzi equations \citep{Eisenhart1925a}.
\begin{eqnarray}
R_{\mu\,\nu\,\lambda\,\delta}&=&2\,e\,b_{\mu[\lambda}\,b_{\delta]\nu},\\
0&=&b_{\mu\,[\nu;\,\lambda]}-\Gamma^{\sigma}_{[\nu\,\lambda]}\,
 b_{\mu\,\sigma}+\Gamma^{\sigma}_{\mu\,[\nu}\, b_{\lambda]\,\sigma},
\end{eqnarray}
where $e=\pm 1$ ($+$ or $-$, when the normal to the manifold is spacelike or timelike respectively) and 
$b_{\mu\,\nu}$ are the coefficient of the second differential form. 
The necessary and sufficient condition for any Riemannian space to 
be an embedding class $\bf{I}$ is the Karmarkar condition \cite{Eisenhart1925b, Karmarkar},
and  for our case, the condition reduces to: 
%
\begin{eqnarray}
R_{rtrt}\,R_{\theta\phi\theta\phi}=R_{r\theta r\theta}\,R_{t\phi t\phi}
-R_{\theta rt\theta}\,R_{\phi rt\phi}.\label{Karmarkar}
\end{eqnarray}
The non vanishing components of the Riemann tensor for the shear free spherically symmetric metric \eqref{1eq1} are 
\begin{eqnarray}
R_{rtrt}&=& a^{2}\left(\frac{a^{''}}{a}-\frac{b^{2}f}{a^{2}}\ddot{f}
-\frac{a^{'}}{a}\frac{b^{'}}{b} \right), \label{Rrtrt}\\
R_{\theta\phi\theta\phi}&=&{r^{4}b^{2}f^{2}}\left(\frac{b^{2}}{a^{2}}\dot{f}^{2}-\frac{2b^{'}}{rb}
-\frac{b^{'\,^{2}}}{b^{2}} \right)\sin^2\theta,\label{Rthphthph}\\
R_{r\theta r\theta}&=&{r^2b^{2}f^{2}}\left(\frac{b^{2}}{a^{2}}\dot{f}^{2}-\frac{b^{'}}{rb}
-\frac{b^{''}}{b}+\frac{b^{'\,^{2}}}{b^{2}} \right),\label{Rrthrth}\\
R_{t\phi t \phi}&=&{r^2a^2b}\left(\frac{a^{'}}{ra}-\frac{b^2f}{a^2}\ddot{f}
+\frac{a^{'}}{a}\frac{b^{'}}{b} \right)\sin^2\theta ,\label{Rtphtph}\\
R_{\theta rt\theta}&=&\frac{r^{2}b^{2}f}{a}a^{'}\dot{f} ,\label{Rthrtth}\\
R_{\phi rt\phi}&=&\sin^2\theta R_{\theta rt\theta} .\label{Rphrtph}
\end{eqnarray}
Using equation \eqref{Rrtrt}-\eqref{Rphrtph} into the equation \eqref{Karmarkar} we have the class {\bf{I}} condition as
\begin{eqnarray}
0&=&b^{2}{\dot{f}}^2b^{3}\left(\frac{a^{''}}{a}-\frac{a^{'}}{ra}+\frac{a^{'\,^2}}{a^{2}}-2\frac{a^{'}}{a}\frac{b'}{b}\right)
+r^{2}b^{3}f\ddot{f}\left(\frac{b^{'}}{rb}+2\frac{b^{'\,^{2}}}{b^{2}}-\frac{b^{''}}{b} \right)\nonumber\\
&+&r^{2}aa^{'}b^{''}\left(\frac{1}{r}+\frac{b^{'}}{b} \right)-r^{2}aba^{''}\left(2\frac{b^{'}}{rb}
+\frac{b^{'\,^{2}}}{b^2}\right)+raba^{'}\left(2\frac{b^{'\,^{2}}}{b^2}+\frac{b^{'}}{rb}\right). \label{Karmarkar1}
\end{eqnarray}
Although we have found the form of one of the metric potential $b(r)$ from the choice of anisotropy factor, 
however, the class {\bf{I}} condition equation \eqref{Karmarkar1} is still nonlinear in its temporal and radial behavior.
For a collapsing model both the conditions \eqref{matching} 
and \eqref{Karmarkar1} must be simultaneously satisfied. It has been found that one of 
the solutions of \eqref{Karmarkar1} is a linear solution \,\cite{Banerjee}
\begin{eqnarray}
f(t)&=&-C_{Z} \,t  \label{f(t)}
\end{eqnarray}
where $C_Z>0$. Using equation \eqref{br} and \eqref{f(t)} into the above equation \eqref{Karmarkar1}
we obtain the form of $a(r)$ as
\begin{eqnarray}
a(r)&=&\frac{1}{2\sqrt{2}\sqrt{C_3C_4}}\sqrt{C_4^{2}\left(C_1b(r)+4C_2C_3\right)^2-4C_Z^2} \label{a(r)}
\end{eqnarray}
where $C_1$ and $C_2$ are integration constants. 
\\
\\
The dependency of gravitational potentials 
$a(r)$ and $b(r)\,f(t)$ in  perfect fluid and the Karmarkar condition leads to two exact solutions:
the Schwarzschild\,\cite{KS} or the Kohler and Chao\,\cite{KC} solution. The Kohler and Chao solution 
is only physical for unbound configuration such as cosmological model
as the radial pressure vanishes at $r\rightarrow \infty$.
The pressure isotropy condition $\Delta=0$ and the Karmarkar condition \eqref{Karmarkar1} are 
satisfied when either $C_{1}, C_{3}$ or $C_{4}$ vanish. However, for the metric potential $a(r)$ to be positive and 
greater than zero, both $C_{3}$ and $C_{4}$ should not vanish throughout the collapse. 
This implies that in the static case, when $f(t)=1$, from pressure isotropy and the Karmarkar condition we 
must have $C_{1}=0$. Thus, the for static case, Karmarkar condition together with the pressure isotropy yields the Schwarzschild like form of the gravitational potentials given by
\begin{eqnarray}
a(r)^{2}&=&\frac{4C_{2}^{2}C_{3}^{2}C_{4}^{2}-1}{2C_{3}C_{4}},\\
b(r)^{2}&=&\frac{4}{\left(2\,C_4+C_3\,r^2\right)^{2}}.
\end{eqnarray}

\begin{figure}[h]
\begin{subfigure}{.5\textwidth}
\centering
\includegraphics[width=\linewidth]{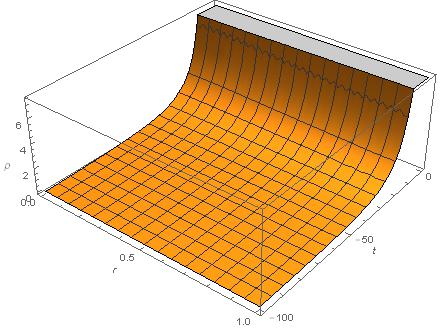}
\caption{}
\label{fig:rho}
\end{subfigure}
\begin{subfigure}{.5\textwidth}
\centering
\includegraphics[width=\linewidth]{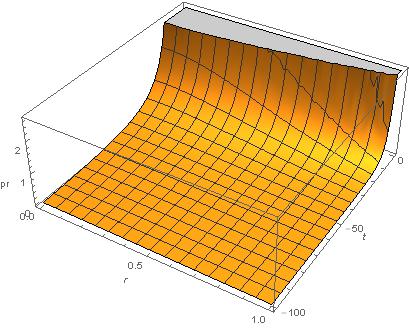}
\caption{}
\label{fig:pr}
\end{subfigure}
\caption{(a) Plot of the density $\rho$, equation \eqref{rhoabf}, with respect to $r$ at center $r=0$ and at the surface of the star at $r=1$. This plot shows that the density is positive as the collapse starts at $t=-100$ and is maximum at the end state of the collapse at $t=0$. (b) Plot of the radial pressure $p_{r}$, equation \eqref{prabf}, with respect to $r$ at center $r=0$ and at the surface of the star at $r=1$. This plot shows that the radial pressure is positive as the collapse starts at $t=-100$ and is maximum at the end state of the collapse at $t=0$. }
\end{figure}
\begin{figure}[h]
\begin{subfigure}{.5\textwidth}
\centering
\includegraphics[width=\linewidth]{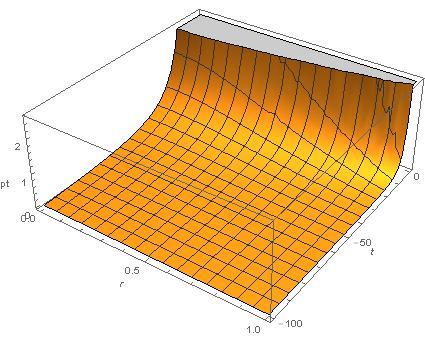}
\caption{}
\label{fig:pt}
\end{subfigure}\begin{subfigure}{0.5\textwidth}
\centering
\includegraphics[width=\linewidth]{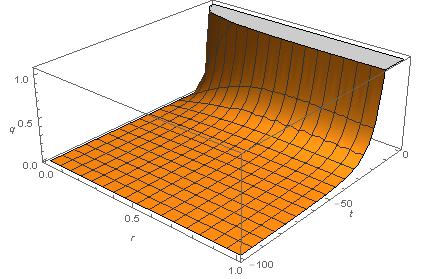}
\caption{}
\label{fig:q}
\end{subfigure}
\caption{ (a) Plot of the tangential pressure $p_{t}$, equation \eqref{ptabf}, with respect to $r$ at center $r=0$ and at the surface of the star at $r=1$. This plot shows that the tangential pressure is also positive as the collapse starts at $t=-100$ and is maximum at the end state of the collapse at $t=0$. (b) Plot of the radial heat flux $q$, equation \eqref{qabf}, with respect to $t$ and $r$. This plot shows that the heat flux is positive, implying heat is radiating throughout the collapse. As the star deviates from the equilibrium, it will start radiating heat and as the contraction of the cloud increases it will start radiating more and more heat flux. It can be seen from the graph that the heat flux starts to increase as the collapse starts at $t=-100$ and heat flux is positive and maximum as it reaches the end state of the collapse at $t=0$.}
\end{figure}
Now physical quantities \eqref{rho}-\eqref{q} 
interms of these exact solutions of the radiating star becomes
\begin{eqnarray}
\rho&=&\frac{6C_{3}C_{4}^{3}}{S_{1}C_{Z}^{2}t^{2}}\left(C_{1}
-2C_{2}C_{3}\left( 2C_{4}+C_{3}r^2\right)\right)^{2},\label{rhoabf}\\
p_{r}&=&\frac{C_{3}C_{4}^{2}}{S_{1}C_{Z}^2t^{2}}\left[2C_{1}C_{2}C_{3}\left(12C_{4}^{2}+4C_{3}C_{4}r^2-C_{3}r^{2}\right)
\right.\nonumber\\
&&-\left.C_{1}^{2}\left(4C_{4}-C_{3}r^2\right)-8C_{2}^{2}C_{3}^2C_{4}\left(2C_{4}+C_{3}r^2\right)^2 \right],\label{prabf}\\
p_{t}&=& \frac{C_{3}C_{4}^{2}}{S_{1}C_{Z}^2t^2}\left[C_{1}^{4} C_{4}^{2} \left(C_{3} r^{2}-4 C_{4}\right)
+2 C_{1}^{3} C_{2} C_{3} C_{4}^2 \left(-3 C_{3}^{2} r^{4}+8 C_{3} C_{4} r^{2}+28 C_{4}^{2}\right)\right.\nonumber\\
&&\left.+2 C_{1}^{2} \left(C_{3} r^{2}+2 C_{4}\right)^{2} \left(6 C_{2}^{2} C_{3}^{2} C_{4}^{2} 
\left(C_{3} r^{2}-6 C_{4}\right)+C_{Z}^{2} \left(2 C_{4}-C_{3} r^{2}\right)\right)\right.\nonumber\\
&&\left.+2 C_{1} C_{2} C_{3} \left(C_{3} r^2+2 C_{4}\right)^{3} \left(C_{Z}^{2} 
\left(C_{3} r^{2}-6 C_{4}\right)-4 C_{2}^{2} C_{3}^{2} C_{4}^{2} \left(C_{3} r^{2}-10C_{4}\right)\right)\right.\nonumber\\
&&\left.-8 C_{2}^{2} C_{3}^{2} C_{4} \left(C_{3} r^{2}+2 C_{4}\right)^4 
\left(4C_{2}^{2} C_{3}^{2}C_{4}^{2}-C_{Z}^{2}\right) \right],\label{ptabf}\\
q&=&-\frac{r C_{1} C_{4}^{5/2} C_{3}^{3/2}}{C_{Z}^{2} t^{3}}\frac{4 \sqrt{2}\left[4 C_{2} C_{3}
-\frac{2 C_{1}}{C_{3} r^{2}+2 C_{4}}\right]}{ \left[C_{4}^{2} 
\left(4 C_{2} C_{3}-\frac{2 C_{1}}{C_{3} r^{2}+2 C_{4}}\right)^{2}-4 C_{Z}^{2}\right]^{3/2}},\label{qabf}
\end{eqnarray}
The boundary condition \eqref{matching} in the view of \eqref{prabf}-\eqref{qabf} becomes
\begin{eqnarray}
2f\,\ddot{f}+\dot{f}^2-2x\dot{f}&=&y ,\label{match}
\end{eqnarray}
where 
\begin{eqnarray}
x&=&\left(\frac{a^{'}}{b} \right)_{\Sigma},\label{x}\\
y&=&\left(\frac{a^2}{b^2}\left[\frac{b^{'\,^{2}}}{b^2}+\frac{2}{r}\left(\frac{b^{'}}{b}
+\frac{a^{'}}{a}\right)+\frac{2a^{'}b^{'}}{ab} \right] \right)_{\Sigma}. \label{y}
\end{eqnarray}
It can be seen from the Figs. \ref{fig:rho}, \ref{fig:pr} and \ref{fig:pt}
that density, radial pressure and tangential pressure are positive throughout the collapse.
Also, Fig. \ref{fig:q} shows that the radial heat flux is also positive throughout collapse.
Also the expansion scalar \eqref{Theta} and the Misner sharp mass function \eqref{mass} have the form
\begin{eqnarray}
\Theta &=& \frac{6\sqrt{C_{3} C_{4}}}{t \sqrt{C_{4}^{2} \left(2 C_{2} C_{3}
-\frac{C_{1}}{C_{3} r^{2}+2 C_{4}}\right)-2 C_{Z}^{2}}},\label{Thetaabf}\\
m&=&\frac{8 t r^{3} C_{3} C_{4}^3 C_{Z}}{\left(C_{3} r^{2}+2 C_{4}\right)^{3}}
\,\left[\frac{2 C_{2} C_{3} \left(C_{3} r^{2}+2 C_{4}\right)-C_{1}}
{2 \left(C_{3} r^{2}+2 C_{4}\right) \left(C_{2} C_{3} C_{4}^{2}-C_{Z}^{2}\right)-C_{1} C_{4}^{2}}\right].\label{mabf}
\end{eqnarray}
\begin{figure}[h]
\begin{subfigure}{.5\textwidth}
\centering
\includegraphics[width=\linewidth]{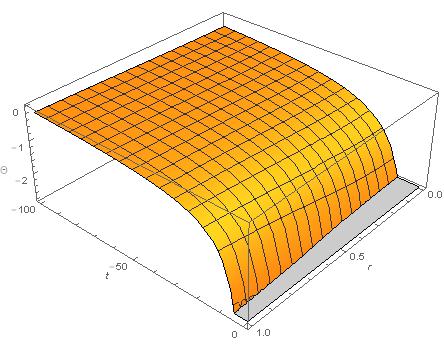}
\caption{}
\label{fig:Theta}
\end{subfigure}\begin{subfigure}{0.5\textwidth}
\centering
\includegraphics[width=\linewidth]{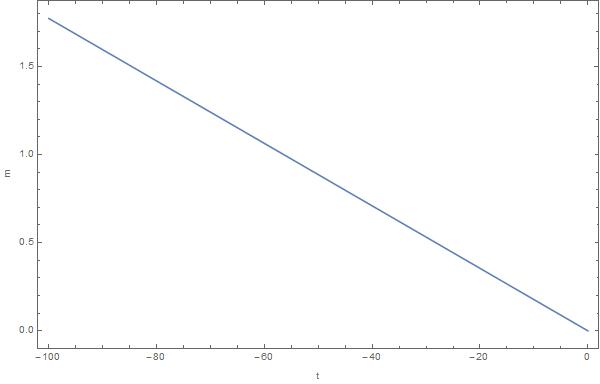}
\caption{}
\label{fig:m}
\end{subfigure}
\caption{ (a) Plot of the expansion scalar $\Theta$, equation \eqref{Thetaabf}, with respect to $t$ and $r$. For the collapsing phenomena $\Theta$ should be negative throughout the collapse which is confirmed from the figure that as the collapse starts at $t=-100$, $\Theta$ has zero value and it starts decreasing till the collapse reaches its end state at $t=0$. (b) Plot of the mass $m$, equation \eqref{mabf} of the star with respect to $t$. It can be seen that the mass is linear function of time coordinate and the mass radiates throughout the collapse.}
\end{figure}
The metric potentials $a(r)$ and $b(r)$ should be greater than zero throughout the collapsing phenomena.
This implies for $b(r)>0$, both $C_{3}$ and $C_{4}$ should be negative. 
For second metric potential to be positive i.e. $a(r)>0$ 
we must have $C_{Z}^2<C_{4}^{2}\left[\frac{C_{1}}{C_{3}r^{2}+2C_{4}}-2C_{2}C_{3}\right]^2$.
This implies at the center $r=0$, $C_{Z}<C_{1}-2C_{2}C_{3}C_{4}$. 
Also, the condition of positivity of the radial pressure $p_{r}$ throughout 
the collapsing phenomena imposes constraints on the range of $C_{1}$ as $C_{2}C_{3}C_{4}<C_{1}<2C_{2}C_{3}C_{4}$. 
\\
\\
Using equations \eqref{br}, \eqref{f(t)}, \eqref{a(r)} and \eqref{prabf} into the \eqref{LInfinity},
we found that the Luminosity of the radiating collapse becomes time independent 
and vanishes if $R^{'}=(b+rb^{'})=0$. Which implies that the red shift diverges 
at the time of formation of the blackhole i.e. at $R^{'}=0$. As we know that $R^{'}$
implies the shell crossing singularity and for a smooth collapse there should be
no shell crossing during the collapse, which means shell crossing singularity should
occur either at the same or later epoch than that of shell focusing singularity( i.e at $R=0$).
As the red shift diverges at $R^{'}=0$, this means the only place where blackhole can form 
is at the central singularity where both the shells  crossing and shell focusing singularity
forms simultaneously. This is because for a physically realistic smooth model there should be
$R^{'}>0$ throughout the collapse. It is clear from the Fig. \ref{fig:Theta} that the 
expansion scalar $\Theta$ is negative throughout implying the contracting behavior of the system.
Also, Fig. \ref{fig:m} shows that the mass is linear function of time $t$.
\\

Recently, the Karmarkar scalar condition has been 
used for the nonstatic system, where they found two class of solutions \cite{ON}. One of 
their solutions is that of a horizon free radiating collapse as in \cite{NGM2018}.
However, the second class of solutions are
\begin{eqnarray}
A(t,r)&=&-\dot{\bar{b}}(t) \frac{\sqrt{C\left(-2+\bar{a}C\right)+\bar{a}r^{4}\left(-1+\bar{a}\,^2 C^2\right)+2r^2\left(-1-\bar{a}C+\bar{a}\,^{2} C^2 \right)}}{{\sqrt{2} (1+\bar{a} r^2)}},\label{A}\\
B(t,r)&=&\frac{\bar{b}(t)}{(1+\bar{a} r^2)}\hspace{0.5cm};\hspace{1cm} R(t,r)=r B(t,r),\label{B}
\end{eqnarray}
It must be noted that, for 
$C_{4}=1,\,C_{Z}=1,\,C_{1}=2C_{Z},\,\bar{b}(t)=f(t)=-C_Z\,t,
\, \bar{a}=C_{3}/2,\, C(t)=-C_{2}$ the
\eqref{A}-\eqref{B} reduce to those obtained here 
using anisotropy together with the Karmarkar condition. 
Thus, although we had chosen $\Delta$ for the mathematical simplicity, 
the results obtained are physically significant and in agreement with those obtained in \,\cite{ON}.
\\

Let us now ask if the solutions also satisfy the energy conditions as well.
Energy conditions plays important role in the study of the astronomical phenomena 
like collapsing stellar models. In this section we will analyze the physical
evidences of our model by verifying the energy conditions. The energy conditions
namely weak energy condition\,(WEC), null energy condition\,(NEC), 
dominant energy condition\,(DEC) and strong energy conditions \,(SEC) 
will be satisfied at all points in the stellar model if the following
inequalities are satisfied simultaneously \cite{Chan1997} \cite{KS}\\
{\bf{E1\,:}}\, $\left(\rho+p_{r}\right)^2-4q^{2}\,\geq \,0$\hspace{3.2cm}(SEC/DEC/WEC)\\
{\bf{E2\,:}}\, $\rho-p_{r}\,\geq\,0\,\,\,\, $\hspace{4.3cm} (DEC)\\
{\bf{E3\,:}}\, $\rho-p_{r}-2p_{t}+\sqrt{\left(\rho+p_{r}\right)^2-4q^{2}}\,\geq\,0 $\hspace{.2cm} (DEC)\\
{\bf{E4\,:}}\, $\rho-p_{r}+\sqrt{\left(\rho+p_{r}\right)^2-4q^{2}}\,\geq\,0\,\,\,\, $\hspace{.9cm} (DEC/WEC)\\
{\bf{E5\,:}}\, $\rho-p_{r}+2p_{t}+\sqrt{\left(\rho+p_{r}\right)^2-4q^{2}}\,\geq\,0$\hspace{.2cm} (SEC/DEC/WEC)\\
{\bf{E6\,:}}\, $2p_{t}+\sqrt{\left(\rho+p_{r}\right)^2-4q^{2}}\,\geq\,0$\hspace{1.7cm} (SEC)\\
Beside these energy conditions, a physically reasonable stellar model should also satisfy \\
{\bf{E7\,:}}\, $\rho>0$,\, $p_{r}>0$,\, $p_{t}>0$, and $\rho^{'}<0$,\, $p_r^{'}<0$,\, $p_{t}^{'}<0$.\\

\begin{figure}[h]
\begin{subfigure}{0.3\textwidth}
\centering
\includegraphics[width=\linewidth]{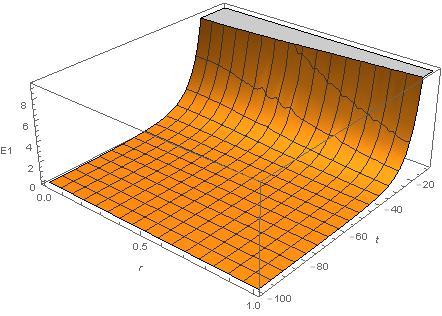}
\caption{}
\label{fig:E1}
\end{subfigure}
\begin{subfigure}{0.3\textwidth}
\centering
\includegraphics[width=\linewidth]{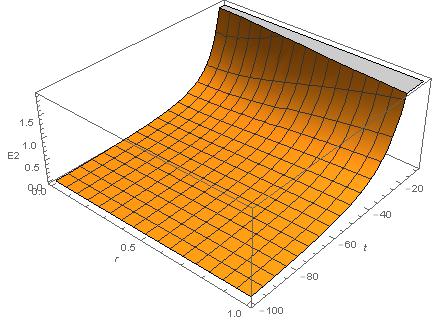}
\caption{}
\label{fig:E2}
\end{subfigure}
\begin{subfigure}{0.3\textwidth}
\centering
\includegraphics[width=\linewidth]{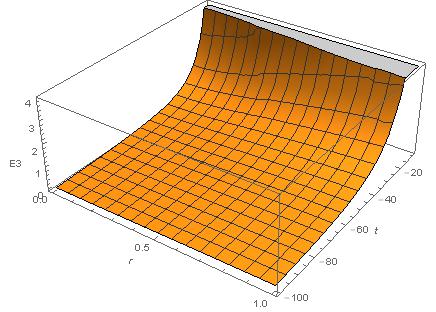}
\caption{}
\label{fig:E3}
\end{subfigure}
\caption{ (a) Plot of the energy condition $E1$ with respect to $t$ and $r$,
(b) Plot of the energy condition $E2$ with respect to $t$ and $r$,
(c) Plot of the energy condition $E3$ with respect to $t$ and $r$.}
\end{figure}

Here, we can see that the validity of the {\bf{E1}} and {\bf{E2}} inequalities implies that
the {\bf{E4}} inequality is satisfied. In the same fashion, the validity of the {\bf{E1}},
{\bf{E2}} and {\bf{E7}} inequalities ensures that the {\bf{E5}} and {\bf{E6}} inequalities
are satisfied. So, in general we only need to check the validity of the 
{\bf{E1}},\, {\bf{E2}},\,{\bf{E3}} and {\bf{E7}}. For our radiating stellar model,
it can be seen from the figures \ref{fig:E1}, \ref{fig:E2} and \ref{fig:E3} 
that all these energy conditions are well satisfied throughout the interior of the collapsing star.

We would like to shed some light on the stability of our model by investigating the shear free condition. 
It has been found that the evolution of the shear depends upon the scalar $Y_{TF}$ which is given by\,\cite{HPO}
\begin{eqnarray}
Y_{TF}&=&\Delta+\frac{\rho}{2}-\frac{3m}{r^3} ,\label{Y_TF}
\end{eqnarray}
where $m$ is Misner Sharp mass function given by \eqref{mass} or interms of the 
density and heat flux it can be written as\,\cite{HPO}
\begin{eqnarray}
m&=&\frac{1}{2}\int^{r}_{0} R^2 \left(R' \rho + \frac{q\,\dot{R}}{a(r)}{b(r)\,f(t)}\right)dr.
\end{eqnarray}
Here, $R=r\,b(r)\,f(t)$ is the radius of the collapsing cloud. Equation \eqref{Y_TF} shows 
that the scalar $Y_{TF}$ depends upon the density of the cloud $\rho$, pressure anisotropy term $\Delta$ and mass function $m$. 
It has been shown that for a geodesic fluid, the presence of the inhomogeneous density $\rho$, 
anisotopy term $\Delta$ and heat flux $q$ can generates the shearing effects within the fluid. 
To check the stability of the shear free condition for our model, we use the forms 
of the pressure anisotropy \eqref{Delta1}, density \eqref{rhoabf}  and mass function
into the equation \eqref{Y_TF}, which gives $Y_{TF}=\Delta$. 
This shows that although we have consider the initial shear free collapse, at later time, 
the presence of the anisotropy term $\Delta$ generates the shearing effects in the fluid distribution.
\\

\section{Thermal properties }\label{sec5}
The study of the thermodynamical evolution and the temperature
profiles of the radiating stars plays prominent role
during dissipative gravitational collapse as they decide the departure from the
thermodynamical equilibrium. Previous studies of the shear free and
shearing dissipative gravitational collapses shows that the relaxation 
effects plays significant role in the temperature profiles towards the end state of the 
dissipative gravitational collapse
see \cite{Martinez1996, Herrera1996, Herrera1997, GMM1999, Herrera2004} and references therein.
To study the thermodynamical evolution and the temperature profiles 
of inside the collapsing star, we will use the causal transport equation for the metric 
\eqref{1eq1}  given by\,\cite{Maartens1995, Martinez1996, Israel1979}
\begin{eqnarray}
\tau h_{\mu}^{\nu}\dot{q}_{\nu}+q_{\mu}&=&-k\left( h_{\mu}^{\nu} \nabla_{\nu} T+T\dot{u}_{\mu} \right)\label{tempgen}\\
\tau \left(qbf\right)_{,t}+q\,a\,b\,f&=&-\frac{k\left(aT \right)_{,r}}{bf}, \label{tempabf}
\end{eqnarray}
where, $\alpha>0$,\,$\beta>0$,\,
$\gamma>0$ and $\sigma>0$ are constants and $h^{\mu\nu}=g^{\mu\nu}+u^{\mu}u^{\nu}$.
Also,
\begin{eqnarray}
 \tau_{c}=\left(\frac{\alpha}{\gamma}\right)\,T^{-\sigma}\hspace{0.2cm},\hspace{0.6cm} 
 k=\gamma\, T^{3}\,\tau_{c}\hspace{0.2cm},\hspace{0.6cm}  \tau=\left(\frac{\beta\,\gamma}{\alpha}\right)\,\tau_{c} \label{tkt}
\end{eqnarray}
 are physically reasonable choices of the mean collision time, between 
massive and massless particles $\tau_{c}$\,,\, thermal conductivity $k$ and 
the relaxation time $\tau$ respectively\,\cite{Martinez1996}\cite{GG2001}. 
$\tau$ represents the causality index, measures the strength of relaxational effects
and $\tau=0$ or $\beta=0$ represents the noncausal case.
Using these forms of the physical quantities given in equation \eqref{tkt},
the form of the causal heat transport equation \eqref{tempabf} becomes
\begin{eqnarray}
\beta T^{-\sigma} \left(qbf\right)_{,t}+q\,a\,b\,f&=&-\frac{\alpha \,\left(aT \right)_{,r}}{bf}\,\,T^{3-\sigma} .\label{tempabff}
\end{eqnarray}
The noncausal solution of heat equations are obtained by setting
$\beta=0$ i.e. $\tau=0$ in the above transport equation \eqref{tempabff}\,\cite{GG2001}
\begin{eqnarray}
\left(a\,T\right)^{4}&=&-\frac{4}{\alpha}\int a^{4}\,q\,b^2\,f^2\,dr+F(t),
\hspace{1.6cm} \sigma= 0  \label{Nsigman4}\\
\ln\left(a\,T\right)&=&-\frac{1}{\alpha}\int q\,b^2\,f^2\,dr+F(t).\hspace{2cm} \sigma=4  \label{Nsigma4}
\end{eqnarray}

The causal solution of the above transport equation \eqref{tempabff} are given by\,\cite{GG2001}
\begin{eqnarray}
\left(a\,T\right)^{4}&=&-\frac{4}{\alpha}\left[\beta\int a^3\,b\,f(q\,b\,f)_{,t}\,dr
+\int a^{4}\,q\,b^2\,f^2\,dr\right]+F(t), \hspace{0.6cm} \sigma=0  \label{Csigman4}\\
\left(a\,T\right)^{4}&=&-\frac{4\beta}{\alpha}exp\left(-\int\frac{4\,q\,b^{2}\,f^{2}}{\alpha}\,dr\right)
\int a^3\,b\,f(q\,b\,f)_{,t}\,dr\,exp\left(\int\frac{4\,q\,b^{2}\,f^{2}}{\alpha}\,dr\right)\nonumber\\
&&+F(t)exp\left(-\int\frac{4\,q\,b^{2}\,f^{2}}{\alpha}\,dr\right),\hspace{4.2cm} \sigma=4 \label{Csigma4}
\end{eqnarray}
where $F(t)$ is the function of integration. The function $F(t)$ is determined by invoking boundary conditions
\begin{eqnarray}
\left(T^{4} \right)_{\Sigma}&=&\left( \frac{L_{\infty}}{4\pi\delta r^{2}b^{2}f^{2}} \right)_{\Sigma}. \label{Tsigma}
\end{eqnarray}
where $L_{\infty}$ is the total luminosity for an observer at infinity and $\delta>0$ is constant. 
 \begin{figure}[h!tbp]
\centering
\includegraphics[scale=0.5,keepaspectratio=true]{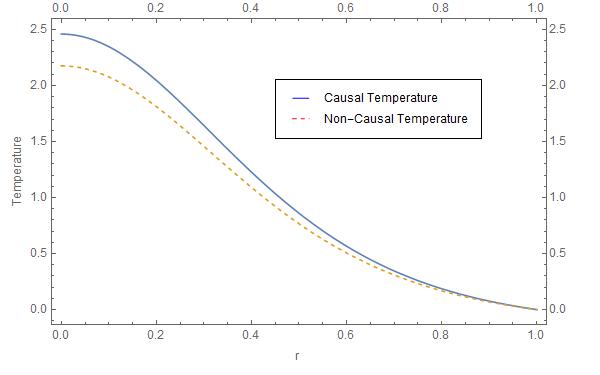}
\caption{Plot of the causal and noncausal temperature $T$ of the star with respect to $r$ for $\sigma=0$.}
\label{fig:Temp}
\end{figure} 
As with previous investigations\,\cite{GMM1998, GG2001}, Fig.\ref{fig:Temp} 
shows that both the causal and noncausal temperature are same at the boundary of the star.
However, at later stages of the collapse, relaxation effects plays significant role 
and they differ from the noncausal case. This behavior can be seen from the 
Fig. \ref{fig:Temp} that with $\beta>0$ the 
relaxations effects grows and the causal temperature remains greater than that 
of noncausal temperature throughout the interior of the star.
These results are in agreement with the earlier results obtained for 
the shear free collapse\,\cite{LHerrera, GMM1999}. 
\section{Conclusion}\label{sec6}
In this paper, we presented new exact solutions of the Einstein field 
equations for spherical symmetric systems with heat-conducting anisotropic fluid as a collapsing matter.
The interior spacetime has been smoothly matched with the exterior Vaidya metric of 
radiant star across the timelike hypersurface $\Sigma$. 
For this purpose, we assume the gravitational potentials to be separated into their radial and temporal coordinates.
Now to find their exact forms, we consider a special form of the pressure anisotropy as given by the equation \eqref{Delta1}
and find one of the metric potential. To find the other gravitational potential,
we employed the Karmarkar condition\,\eqref{Karmarkar1}, which makes the interior spacetime to be class {\bf{I}}.

We have investigated the physical quantities like density\,\eqref{rhoabf}, radial pressure\,\eqref{prabf} and 
tangential pressure\,\eqref{ptabf} and it can be seen from the Figs.\,\ref{fig:rho}, \ref{fig:pr} and \ref{fig:pt}
that they are positive throughout the collapse. From Fig.\,\ref{fig:q} 
it is clear that the radial heat flux\,\eqref{qabf} is finite and positive throughout collapse. 
From Fig.\,\ref{fig:Theta}, the negative form of the expansion scalar\,\eqref{Thetaabf} 
implies the collapsing behavior of the system.
We have also investigated the mass function and total luminosity of the collapsing star.
It is clear from the Fig.\,\ref{fig:m} that the mass\,\eqref{mabf} depends linearly on the temporal coordinate. 
It has been found that the luminosity is time independent for this class of solutions 
and it radiates uniformly throughout the collapse. Thus the solution is a physically permissible solution
of the Einstein theory. We also checked the physically viability  of the collapsing model by studying the 
energy conditions {\bf{E1}}, {\bf{E2}} and {\bf{E3}} together with the positivity 
of the density, radial and tangential pressure profiles and heat flux.
It is clear from the figures \ref{fig:E1},\,\ref{fig:E2} and\,\ref{fig:E3} that all 
three energy conditions {\bf{E1}}, {\bf{E2}} and {\bf{E3}} are satisfied. 
It shows that the present class of solutions seams to be representing the physically viable collapse model.

Also, we have found the surface temperature of the collapsing star at a large past time.
It is clear from the Fig.\,\ref{fig:Temp} that both the causal and noncausal temperatures 
are same at the boundary $\Sigma$, however, differs at all interior points of collapsing star. 
At later stages of the collapse, relaxation effects\,\eqref{Csigman4} plays significant role and
remains greater than that of the noncausal case\,\eqref{Nsigman4}.
From Fig.\,\ref{fig:Temp} shows that with $\beta>0$ the 
relaxations effects grows and the causal temperature remains greater than that 
of noncausal temperature throughout the interior of the star.
These results are in agreement with the earlier results obtained for 
the shear free collapse.
\\\\{\bf{Acknowledgments}}\\
The author thanks Dr. Ayan Chatterjee for several stimulating
discussions and continuous encouragement during the course of the work.
He also acknowledges his suggestions on various aspects of the work that made the presentation better.
The author also acknowledges the comments of the referees which led to several improvements.

\end{document}